\documentclass[12pt]{article}

\usepackage[body={17cm, 23cm, centered}]{geometry}
\usepackage{epsfig,verbatim,cite}
\usepackage{lmodern}
\usepackage{braket}
\usepackage{lmodern}
\usepackage[utf8]{inputenc}			

\usepackage{mathtext,marvosym,textcomp}
\usepackage{mathtools}
\usepackage{slashed}
\usepackage[dvipsnames,svgnames,table]{xcolor}

\usepackage{tikz}
\usetikzlibrary{arrows.meta}

\usepackage[ normalem]{ulem}
\usepackage{tocloft}
\usepackage{epsfig,amsmath,amsfonts,amssymb,arydshln}
\usepackage[makeroom]{cancel}
\usepackage{multirow}
\usepackage{aligned-overset}
 
 \usepackage[debug,pageanchor=false]{hyperref}
\hypersetup{colorlinks=true,linktocpage,breaklinks,
            urlcolor=blue,
            linkcolor=red,
            citecolor=blue
            }

\numberwithin{equation}{section}

\def\a{\alpha} 
\def\b{\beta} 
\def\g{\gamma} 
\def\d{\delta} 
\def\e{\epsilon}
\def\ve{\varepsilon} 
 
\def\h{\eta} 

\def\k{\kappa} 
\def\l{\lambda} 
\def\m{\mu}
\def\n{\nu} 
\def\x{\xi} 
\def\p{\pi}
\def\r{\rho}
\def\q{\theta}
\def\s{\sigma} 
\def\t{\tau}  
\def\f{\phi}

\def\w{\omega}

\def\G{\Gamma}

\def\S{\Sigma}

\def\W{\Omega}

\def\fr{\frac}

\def\dt{\partial}

\def\ph{\phantom}

\def\mc{\mathcal}

\def\mF{\mathcal{F}}

\def\td{\tilde{\d}}
\def\tG{\tilde{\G}}

\def\tG{\tilde{G}}

\def\SS{\mathbb{S}}

\def\AdS{\mathrm{AdS}}

\def\rmU{\mathrm{U}}

\def\rmE{\mathrm{E}}

\def\vol{\text{vol}}

\def\diag{\mathrm{diag}}

\usepackage{fontawesome5}

\usepackage{cmap}					

\usepackage[pdftex]{pict2e}
\usepackage[dvipsnames]{xcolor}

\usepackage{tcolorbox}

\begin{document}

\begin{titlepage}
\ph{preprint}

\vfill

\begin{center}
   \baselineskip=16pt
   {\large \bf Brane bound states, deformations and OM.  
   }
   \vskip 1cm
    Sergei Barakin$^{a,d}$\footnote{\tt barakin.serge@gmail.com },
    Kirill Gubarev$^{b,d,a}$\footnote{\tt kirill.gubarev@phystech.edu }
    and Edvard T. Musaev$^{c,d}$\footnote{\tt emusaev@theor.jinr.ru}
       \vskip .6cm
            \begin{small}
                          {\it
                          $^a$Institute of Theoretical and Mathematical Physics, Moscow State University, 119991, Russia
                          $^b$Institute for Information Transmission Problems, 127051, Moscow, Russia\\
                          $^c$Bogoliubov Laboratory of Theoretical Physics, Joint Institute for Nuclear Research, \\ 6, Joliot Curie, 141980 Dubna, Russia\\
                          $^d$Moscow Institute of Physics and Technology, 
                          Laboratory of High Energy Physics, \\
                          9, Institutskii pereulok, 141702, Dolgoprudny, Russia
                          } \\ 
\end{small}
\end{center}

\vfill 
\begin{center} 
\textbf{Abstract}
\end{center} 
\begin{quote}
We investigate behaviour of brane backgrounds under poly-vector deformations in Type IIB and D=11 supergravities. We find that the standard bi-vector deformations add dissolved F1 string charge to a Dp-brane background, quadri-vector deformations add D3-brane charge, tri- and six-vector deformations in D=11 add M2- and M5-brane charges respectively. We discuss these results in the context of NRCS and OM theories. 
\end{quote} 

\vfill
\setcounter{footnote}{0}
\end{titlepage}

\tableofcontents

\setcounter{page}{2}

\section{Introduction}

Bi-vector deformations, understood as transformations of a string background mapping supergravity solutions into solutions, allow to travel between sectors of string theory with very different properties. Probably the most well-known example of this is given by the non-commutative deformations of gauge theories living of Dp-branes first observed in \cite{Arfaei:1997hb,Sheikh-Jabbari:1997qke} as a non-commutativity of brane bound states for various boundary conditions. Later in \cite{Seiberg:1999vs} it was shown that effectively dynamics of open strings on a background given by the metric $g$ and the B-field (that correspond to coherent states of closed strings) is described in terms of a different metric $G$ and the so-called non-commutativity parameter $\q$. The relation between these two sets of fields is given by the so-called Seiberg--Witten map
\begin{equation}
\label{eq:SW_map}
    G^{-1} + \fr{1}{2\p \a'}\q = (g+ 2\p \a' B)^{-1}.
\end{equation}
The metric $G$ is commonly referred to as the open string metric, hence, the map is often called the open-closed map. More concretely, consider open string in a B-field background
\begin{equation}
    S_{OS} = \fr{1}{4 \pi \a'} \int_\S d^2 \s \Big(g_{\m\n}\h^{ab} + 2\p \a' B_{\m\n}\e^{ab}\Big)\dt_a x^\m \dt_b x^\n - \int_{\dt \S}A_\m \dt_\t x^\m.
\end{equation}
In the limit of slowly varying fields, that can roughly be formulated as $\sqrt{2\p \a'}|\dt F/F|\ll 1$ the effective action governing dynamics of the open string ends is given by the DBI action \cite{Fradkin:1985qd,Tseytlin:1999dj}
\begin{equation}
\label{eq:DBI}
    S_{DBI} = \fr{2\p}{g_s (4\p^2 \a')^{\fr{p+1}{2}}}\int d^{p+1}\x\sqrt{-\det \Big(g+ 2\p \a' (B+F)\Big)}.
\end{equation}
Here $F=dA$ is the world-volume field strength. The Seiberg--Witten map states that  this action is equivalent to
\begin{equation}
\label{eq:DBI_NC}
    \hat{S}_{NC} = \fr{2\p}{G_s (4\p^2 \a')^{\fr{p+1}{2}}}\int d^{p+1}\x \sqrt{-\det \Big(G+2\p\a' \hat{F}\Big)},
\end{equation}
where $\hat{F} = d \hat{A} - [\hat{A},\hat{A}]_*$ is the non-commutative field strength and $*$ is the standard non-commutative Moyal product defined by $\q$. The open string metric is given by the SW map and the coupling parameter $G_s$ is defined as
\begin{equation}
    G_s = g_s\sqrt{\fr{\det G}{\det (g+2\p\a' B)}}.
\end{equation}
In the zero-slope limit $\a'\to 0$ when the string oscillatory modes decouple the DBI action \eqref{eq:DBI} with vanishing B-field gives the standard Yang--Mills action. The same limit is well defined in the case when $B$ has only spatial components, in which case \eqref{eq:DBI_NC} gives the action of the non-commutative Yang--Mills theory.  Formally the limit is taken by rescaling
\begin{equation}
    \begin{aligned}
        \a' & = \e^{\fr12}\bar{\a}', && g_{ij} = \e \bar{g}_{ij},
    \end{aligned}
\end{equation}
where barred quantities are the effective parameters of the theory after sending $\e \to 0$. The non-commutativity parameter and the open string metric in the directions where $B\neq 0$ then become
\begin{equation}
    \begin{aligned}
        \q &  \to -B^{-1}, \\
        G^{-1} &\to -(2\p \a')^{-1} B^{-1}g B^{-1}.
    \end{aligned}
\end{equation}

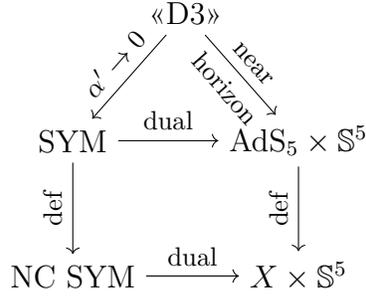
\begin{figure}
    \centering
    \begin{tikzpicture}
        \node at (0,0)    (D3) {<<D3>>};
        \node at (-1.5,-1.7)  (YM) {SYM};
        \node at (1.5,-1.7)  (AdS) {$\AdS_5\times \SS^5$};
        \node at (-1.5,-3.5)  (NCYM) {NC SYM};
        \node at (1.5,-3.5)  (def) {$X\times \SS^5$};
        \draw[->] (D3) -- (YM) node [pos=.5, above, sloped]{\footnotesize$\a' \to 0$};
        \draw[->] (D3) -- (AdS) node [pos=.5, above, sloped]{\footnotesize{near}} node [pos=.5, below, sloped]{\footnotesize{horizon}};
        \draw[->] (YM) -- (AdS) node [pos=.5, above, sloped]{\footnotesize{dual}} ;
        \draw[->] (NCYM) -- (def) node [pos=.5, above, sloped]{\footnotesize{dual}} ;
        \draw[->] (YM) -- (NCYM) node [pos=.5, above, sloped, rotate=180]{\footnotesize{def}} ;
        \draw[->] (AdS) -- (def) node [pos=.5, above, sloped, rotate=180]{\footnotesize{def}} ;
    \end{tikzpicture}
    \caption{Relations between near-horizon limits, dual gauge theories and their deformations. <<D3>> here denotes DBI action or the corresponding supergravity background, and $X$ denotes the deformed $\AdS_5$.}
    \label{fig:NC}
\end{figure}

Translation of this picture into the language of bi-vector deformations is straightforward. In the zero-slope limit, that can be made equivalent to the low energy regime, closed string dynamics near the Dp-brane and far from it decouple. This observation stands behind the AdS/CFT correspondence as suggested in \cite{Aharony:1999ti}, that states that near the horizon of the Dp-brane closed and open string describe the same physics. 
The renowned example is the equivalence of closed string theory on the $\AdS_5\times \SS^5$ background at small coupling and $D=4$ $\mc{N}=4$ SYM at large coupling. Now, the map \eqref{eq:SW_map} can be understood as a transformation from a closed string background given by the metric $g(\q=0)=G$ to the background given by the metric $g$ and the Kalb--Ramond field $B$. Analysis of the symmetry algebra shows that taking $\q$ non-vanishing in AdS directions this map indeed gives a background dual to the non-commutative Yang--Mills theory \cite{Lunin:2005jy,vanTongeren:2015uha,Araujo:2017jkb,Araujo:2017jap}. This can be illustrated by a diagram as in the Fig. \ref{fig:NC}

When the bi-vector  can be represented as $\q=\dt_{\f_1}\wedge \dt_{\f_2}$, where $\f_1$ and $\f_2$ are two (angular) isometry directions, the deformation is equivalent to performing T-duality along $\f_1$, a shift $\f_1 \to \f_1 + \b \f_2 $ and again T-duality along $\f_1$. Here $\b$ is a constant deformation parameter. Behaviour of Dp-branes under such TsT transformation is pretty well understood and summarized in \cite{Imeroni:2008cr}. When the  bi-vector $\q$ has components along time-like directions (equivalently, the same for the B-field) the brane physics becomes essentially different. To start with one is not able to perform TsT transformation as these would involve time-like T-dualities. A way to override this is to use bi-vectors with non-vanishing components along light-like directions. Analysis of the near-horizon symmetries of such deformed background reveals the Schroedinger algebra, that is the non-relativistic counterpart of the (super-)conformal algebra \cite{Imeroni:2009cs}.

Non-relativistic nature of such backgrounds is consistent with the analysis of the closed \cite{Gomis:2000bd} and open \cite{Seiberg:2000ms} string spectrum with space-time non-commutativity presented earlier. The main point of these constructions is that it is not possible to consistently take the zero-slope limit in this case. Instead, one has to simultaneously take the space-time components of the B-field to their critical value and rescale $\a'$. Let us illustrate this on a simple example when only one component, say $B_{01}=E$, is non-zero. Taking the closed string metric diagonal 
\begin{equation}
    ds^2 = \g (-(dx^0)^2 + (dx^1)^2) + \d dx_{\perp}^2,
\end{equation}
we obtain the following scaling for the open string metric and the non-commutativity parameter in the $\{01\}$-block:
\begin{equation}
    \begin{aligned}
        G& = \big(\e^2 \g^2 - \e (2\p \a')^2 E^2\big)e^{-1} \g^{-1}
            \begin{bmatrix}
                -1 & 0 \\
                 0 & 1
            \end{bmatrix},\\
        \q & = \fr{(2\p \a')^2 E \e}{\e^2 \g^2 - \e (2\p \a')^2 E^2}
            \begin{bmatrix}
                0 & -1 \\
                1 & 0
            \end{bmatrix}.
    \end{aligned}
\end{equation}
Now we see that upon $\e \to 0$ the metric $G$  degenerates and $\q$ diverges as $E$ approaches a critical value, that is $E_{cr}=\g/2\p\a'$ in this case. In \cite{Seiberg:2000ms} it has been shown, that when $E \to E_{cr}$ the open string dynamics becomes essentially different, in particular its ends become separated by a light-like interval and no longer interact. The open string spectrum becomes non-relativistic and the same is true for the closed string on a particular background studied in \cite{Gomis:2000bd}. To arrive at this Gomis--Ooguri background the proper zero-slope limit must be taken as follows
\begin{equation}
    \begin{aligned}
        \fr{\a'}{\a'_{eff}}&=\e\to 0, \\
        1 - \left(\fr{2\p\a' E}{\g}\right)^2 & = \e \to 0,\\
        2 \p \a' \G^{-1} & = 2 \p \a'_{eff}\G_{eff}^{-1},
    \end{aligned}
\end{equation}
where we use the notations $G = \G \,\diag[-1,1]$. The first line above defies the effective string tension, the second line corresponds to taking $E \to E_{cr}$ and the third line is required to have a finite non-commutativity parameter. In this limit it becomes
\begin{equation}
    \q \to 2 \p \a'_{eff} \tG^{-1}
        \begin{bmatrix}
            0 & -1 \\
            1 & 0
        \end{bmatrix}.
\end{equation}
To keep all blocks of the open string metric of the same order one has to rescale the remaining components of the metric $    \d = \td \e$. The action then rescales as follows
\begin{equation}
\label{eq:action}
    \begin{aligned}
        S & = \fr{1}{4\p\a'}\int d^2 \s \Big[\g (\dt x_{||})^2 + \d (\dt_\perp)^2- 2 (2\p\a')E \dt x^0 \dt x^1\Big] \\
         & = \fr{1}{4\p\a'_{eff}}\int d^2 \s \Big[\g \e^{-1}(\dt x_{||})^2 +  \td (\dt x_\perp)^2- 2 \g \left(\fr{1}{\e}-\fr{1}{2}\right) \dt x^0 \dt x^1\Big].
    \end{aligned}
\end{equation}
Collecting all rescalings together the closed string background can be written as follows
\begin{equation}
    \begin{aligned}
         ds^2 & = \g (-(dx^0)^2 + (dx^1)^2) + \e \td dx_{\perp}^2, \\
         2\p \a' B & = \g \left(1 - \fr{\e}{2}\right).
    \end{aligned}
\end{equation}
In this form the non-relativistic limit of closed string background has been considered in \cite{Gomis:2000bd}. The effective closed string background can be read off from the second line of \eqref{eq:action} and written as
\begin{equation}
    \begin{aligned}
         ds^2 & = \e^{-1} \g (-(dx^0)^2 + (dx^1)^2) + \td dx_{\perp}^2, \\
         2\p \a' B & = \g \left(\fr{1}{\e} - \fr{1}{2}\right)dx^0\wedge dx^1.
    \end{aligned}
\end{equation}
This background with the notation $\w^2 = \e^{-1}$ is the same as the one used in \cite{Blair:2020ops} to interpret taking the non-relativistic limit as moving along a family of non-degenerate generalized metrics.

The main difference between the cases of space and space-time non-commutativity (magnetic and electric deformations respectively) is that in the former case one is able to recover a field-theoretical description, while in the latter case the dynamics remains essentially stringy. The theory that contains open strings ending on Dp-branes with critical electric field and closed strings on the corresponding backgrounds is referred to as the non-relativistic string theory (NRST). For historical reasons its open string sector is referred to as the non-commutative open strings (NCOS). Due to the absence of the field theory limit the question of what is the electric deformation of, say, the $D=4$ $N=4$ SYM is ill-posed. Instead one should ask about the dual holographic description (in certain sense) of the whole NRST, that has been suggested in \cite{Harmark:2000wv}. The idea is that similarly to how Dp-brane backgrounds give low energy closed string description of the effective open string backgrounds of Dp-F1 bound states give low energy non-relativistic closed string (NRCS) description of NCOS. In particular, using this conjecture it has been shown that the Hagedorn phase transition never occurs in NRST in consistency with the earlier results of \cite{Barbon:2001tm}. 

In this work we analyse the behaviour of Dp-brane backgrounds under bi-vector deformations and under their generalization quadri-vector deformations,  Investigating the examples of the D1 and D3 branes we show that deforming a Dp-brane background by a deformation with bi-vector $\b_2=\b \dt_0\wedge\dt_1$ produces backgrounds of Dp-F1 bound states. We show that the standard bi-vector deformation gives precisely the Dp-F1 bound state backgrounds of \cite{Harmark:2000wv}. Its S-dual bi-vector gives D3-D1 bound states as expected and the quadri-vector deformation adds the D3-brane charge. Applied to F1, D1 and D3 branes the bi-vector, its S-dual and quadri-vector deformations respectively change the core charge of the background as 
\begin{eqnarray}
    Q \to \fr{Q}{1 - \b^2},
\end{eqnarray}
where $\b$ is the deformation parameter. We call this process the sedimentation of branes since the deformation acts formally in the while space-time, while the charge changes eventually at the origin.

The same sedimentation we observe for 11-dimensional tri-vector deformations applied to the M2-brane backgrounds. Applied to the M5-brane background a tri-vector deformation produces the M5-M2 brane bound state of \cite{Harmark:2000ff}. In this work such a bound state background was considered as the low energy description of space-time <<non-commutative>> open membranes in the critical C-field limit first considered in \cite{Gopakumar:2000ep,Bergshoeff:2000ai}  and dubbed <<the OM theory>> in \cite{Gopakumar:2000ep}. We show that the tri-vector deformations are in the same relation to the open-closed membrane map of \cite{Bergshoeff:2000jn} as the bi-vector deformations are to the Seiberg--Witten map \eqref{eq:SW_map}. Such a direct relation and a complete analogy with the 10D case provides further confirmation for the M5-M2 bound state background to be understood as low energy description of the OM theory. We also show that the same is true for the open Dp-brane metrics and generalized non-commutative parameters of \cite{Berman:2001rka}. This is consistent with the relation analysed in \cite{Russo:2000zb,Russo:2000mg} between reductions of the OM theory and NCOS.

\section{Branes and strings: NRST}

In this Section we consider deformations of 10-dimensional Type II supergravity backgrounds and their deformations by the standard bi-vector as in \eqref{eq:SW_map}, its S-dual and quadri-vector. Notations for bi-vector deformations of the NS-NS and R-R fields are summarized in Appendix \ref{app:bi}. Quadri-vector deformations will be performed along the rules derived in \cite{Gubarev:2024tks} and summarized in Appexndic \ref{app:quadri}. These essentially require breaking of the D=10 symmetry to combine fields into U-duality multiples. 

A general background generated by a stack of $N$ extremal Dp-branes is given by
\begin{equation}
\label{eq:D3_initial}
    \begin{aligned}
        ds^2 &= H_0(r)^{-\fr12}dy_{||}^2 + H_0(r)^{\fr12}\big(dr^2 + r^2 d\W_{8-p}\big), \\
        \mc{F}_{01\dots p+1} &= -(-1)^p H_0(r)^{-2}\dt_r H_0(r), \\
        e^{2\f} & = e^{2\f_0} H_0(r)^{\fr{3-p}{2}},
    \end{aligned}
\end{equation}
where the harmonic function and the dilaton $\f_0$ are defined in terms of the string length $l_s = \sqrt{\a'}$ and the string coupling constants $g_s$ as follows
\begin{equation}
    \begin{aligned}
        H_0(r)& = 1 + \fr{r_0^{7-p}}{r^{7-p}}, \quad r_0^{7-p} = \fr{ g_s (2\p l_s)^{7-p}}{(7-p)\w_{(8-p)}}N, \\
        e^{\f_0} & = g_s.
    \end{aligned}
\end{equation}
The numerical parameter $\w_{(n)}$ is the volume of the unit $n$-dimensional sphere and reads
\begin{equation}
    \w_{(n)} = \fr{2\p^{\fr{n+1}{2}}}{\G(\fr{n+1}{2})}.
\end{equation}
Below we consider deformations of a stack of D1 and of D3 branes. The harmonic function $H_0(r)$ will change and the coordinates $y_{||}$ will be rescaled, hence the notations.

\subsection{D3-F1 bound state}
\label{sec:D3F1}

Let us start with the standard bi-vector deformation  with the bi-vector given by $\b \dt_0\wedge \dt_1$, where $\b=$const and consider the case of the extremal D3-brane background. For the deformed solution we obtain
\begin{equation}
    \begin{aligned}
        ds^2 =&\  H_0^{\fr12}\big(H_0 - \b^2\big)^{-1} \big[-(dy^0){}^2 + (dy^1){}^2 \big]+ H_0^{-\fr12}\big[ (dx^2){}^2 +  (dx^3){}^2\big] \\
          &+  H_0^{\fr12} \big[ dr{}^2 + r^2 d\W^2\big],\\
         \mF_5 =&\ H_0^{-1}(H_0 - \b^2)^{-1}\dt_r H_0\, dy^0\wedge dy^1\wedge dy^2 \wedge dy^3 \wedge dr \\
         \mF_3 = & \b H_0^{-2}\dt_r H_0 \,dy^2 \wedge dy^3 \wedge dr \\
         B =&\ \b \big(H_0 - \b^2\big)^{-1} dy^0\wedge dy^1\\
         e^{2\Phi} = &\ e^{2\f_0} H_0 \big(H_0 - \b^2\big)^{-1},
    \end{aligned}
\end{equation}
where we denote $y^{2,3}=x^{2,3}$. To show that this is precisely the D3-F1 bound state background analysed in \cite{Harmark:2000wv} one has to rescale the coordinates as 
\begin{equation}
    \begin{aligned}
        x^{0,1} &= (1-\b^2)^{-1/2}y^{0,1}
    \end{aligned}
\end{equation}
and define
\begin{equation}
    \begin{aligned}
        H & = (1-\b^2)^{-1}\left(H_0 - \b^2\right) = 1 + \fr{r_0^3}{1-\b^2}\fr{1}{r^2}, \\
        D^{-1}& = (1-\b^2)\left(1 - H_0^{-1} \b^2\right)^{-1} = \cos^2 \q + \sin^2 \q H^{-1},\\
        e^{2{\Phi_0}} & =(1-\b^2) e^{2{\f}_0}
    \end{aligned}
\end{equation}
Where the parameter $\q$ has been introduced as $\b = \sin \q$. We see that the core charge gets rescaled, that will also be the case of sedimentation. The background is then rewritten as
\begin{equation}
\label{eq:extrD3F1}
    \begin{aligned}
        ds^2 =&\ D^{-\fr12} H^{-\fr12} \big[- (dx^0){}^2 + (dx^1){}^2\big]+D^{\fr12} H^{-\fr12} \big[(dx^2){}^2 + (dx^3){}^2\big] \\
        &+ D^{-\fr12} H^{\fr12} \big[ dr{}^2 + r^2 d\W^2\big], \\
        e^{2\Phi}  =&\ e^{2\bar{\Phi}_0} D^{-1}, \\
        H_3 =&\ \sin \q \, \dt_r H^{-1} dx^0\wedge dx^1 \wedge dr, \\
        \mF_5 =&\ D H^{-2}\dt_r \left(D^{-1}H\right)\, dx^0\wedge dx^1\wedge dx^2 \wedge dx^3 \wedge dr \\
         \mF_3 = & -\sin \q \, \dt_r \left(D^{-1}H\right) \,dx^2 \wedge dx^3 \wedge dr ,
    \end{aligned}
\end{equation}
and reproduces the D3-F1 background used in \cite{Harmark:2000wv}\footnote{We correct here a few misprints in the R-R fluxes occurred in the text of \cite{Harmark:2000wv}.} As expected in the case of an electric deformation one observes a critical value of the deformation parameter, that in the units chosen here is simply $\b^2 =1$. It is natural to expect that taking $\b^2 \to 1$ together with the appropriate rescaling of the coordinates parallel to the brane world-volume gives the same non-relativistic near-horizon limit as in \cite{Harmark:2000wv}. 

The calculation presented above allows to conclude that a bi-vector deformation of a D3 (more generally, any Dp)-brane background gives a D3(Dp)-F1 bound state. The string is dissolved in the D3-brane world-volume and hence there is no limit that allows to turn off the D3 to get a stand-alone F1. Since bi-vector deformations are conceptually only a special case of more general poly-vector deformations one naturally expects that the same picture must hold in general. More concretely: in Type II theory polyvectors add dissolved Dq-branes to a Dp brane, in 11D theory tri-vectors add dissolved M2-branes, six-vectors add dissolved M5-branes. It is then natural to consider the resulting M5-M2 bound state background as a supergravity description of the OM theory in the same sense as the D3-F1 bound state background describes NRST. The rest of the paper is devoted to providing more examples supporting and enriching this concept.

\subsection{D1-F1 bound state}
\label{sec:D1F1}

Another example to consider is the bi-vector deformation of the D1-brane background. The resulting solution reads
\begin{equation}
    \begin{aligned}
        ds^2 =&\  H_0^{\fr12}\big(H_0 - \b^2\big)^{-1} \big[-(dy^0){}^2 + (dy^1){}^2 \big]+  H_0^{\fr12} \big[ dr{}^2 + r^2 d\W^2\big],\\
         \mF_3 =&\ H_0^{-1}(H_0 - \b^2)^{-1}\dt_r H_0\, dy^0\wedge dy^1 \wedge dr \\
         \mF_1 = & -\b H_0^{-2}\dt_r H_0 \, dr \\
         B =&\ \b \big(H_0 - \b^2\big)^{-1} dy^0\wedge dy^1\\
         e^{2\Phi} = &\ e^{2\f_0} H_0^2 \big(H_0 - \b^2\big)^{-1},
    \end{aligned}
\end{equation}
where now
\begin{equation}
    H_0= 1 + \fr{r_0^6}{r^6}.
\end{equation}
To arrive at the solution in the form used in \cite{Harmark:2000wv} one has to rescale the coordinates in the same way as before and to define the functions $D$ and $H$. All the expressions completely repeat those of the previous section.

\subsection{Thermal D3-F1}

The backgrounds analysed in \cite{Harmark:2000wv} are non-extremal and correspond to Dp-F1 states at finite temperature. It is natural to ask, whether a bi-vector deformation of the non-extremal Dp-brane background gives the thermal Dp-F1 bound state. The answer appears to be <<no>>. Let us give more details though focusing at the case of the non-extremal D3-brane whose  background solution reads
\begin{equation}
\label{eq:D3-F1_t}
    \begin{aligned}
        ds^2 =&\  H_0^{-\fr12} \big[-f (dy^0){}^2 + (dy^1){}^2 \big]+ H_0^{-\fr12}\big[ (dx^2){}^2 +  (dx^3){}^2\big] \\
          &+  H_0^{\fr12} \big[ f^{-1}dr{}^2 + r^2 d\W^2\big],\\
         \mF_5 =&\ \a H_0^{-2}\dt_r H_0\, dy^0\wedge dy^1\wedge dy^2 \wedge dy^3 \wedge dr \\
         e^{2\phi} = &\ e^{2\f_0} ,
    \end{aligned}
\end{equation}
where the $f(r)$ is the blackening function defined as 
\begin{equation}
    f = 1 + \fr{h}{r^4}(1 - \a^2).
\end{equation}
The parameter $\a$ is related to the temperature of the state.

The bi-vector deformed background is given by 
\begin{equation}
\label{eq:D3-F1_t_def}
    \begin{aligned}
        ds^2 =&\  H_0^{\fr12}\big(H_0 - \b^2 f\big)^{-1} \big[-f (dy^0){}^2 + (dy^1){}^2 \big]+ H_0^{-\fr12}\big[ (dx^2){}^2 +  (dx^3){}^2\big] \\
          &+  H_0^{\fr12} \big[ f^{-1}dr{}^2 + r^2 d\W^2\big],\\
         \mF_5 =&\ \a H_0^{-1}(H_0 - \b^2 f )^{-1}\dt_r H_0\, dy^0\wedge dy^1\wedge dy^2 \wedge dy^3 \wedge dr \\
         \mF_3 = & \a \b H_0^{-2}\dt_r H_0 \,dy^2 \wedge dy^3 \wedge dr \\
         B =&\ \b \big(H_0 - \b^2 f\big)^{-1} dy^0\wedge dy^1\\
         e^{2\Phi} = &\ e^{2\f_0} H_0 \big(H_0 - \b^2 f \big)^{-1}.
    \end{aligned}
\end{equation}
Form of the gauge fields requires rescaling of the coordinates precisely as in the extremal case, while comparison to the metric of the solutions in \cite{Harmark:2000wv} gives
\begin{equation}
    D^{-1} = \cos^2\q + \sin^2\q H^{-1} f(r),
\end{equation}
i.e. an additional factor $f(r)$. The difference between the derived solution and the thermal Dp-F1 background of \cite{Harmark:2000wv} may be explained in several ways. First, there is a subtlety in performing T-duality along the Euclidean compact time direction. This may affect the interpretation of the result as a proper string background. Second: a bi-vector deformation might add only extremal dissolved strings that are not of <<the same temperature>> as the initial thermal D3-brane background. In this case one should observe some sort of instability reflecting thermal non-stationarity of the state. Finally, the boldest statement would be that near-horizon limit of such obtained solution provides the correct description of thermal NRST. At this point it is not clear to us how to distinguish between all these possibilities, therefore we only mention them here and reserve more detailed investigation for a future work. We give however a hint towards the second version in the following Section.

\subsection{Sedimentation of the string and Dp-branes}

\textbf{F1}

The bi-vector deformation considered in previous sections generated the Dp-F1 bound state when applied to the extremal Dp-brane background. We show here that this is no coincidence and such a deformation indeed adds the dissolved string charge. For that consider the extremal F1-string solution to Type II supergravity equations:
\begin{equation} 
\label{eq:F1_initial}
\begin{aligned}
    ds^2 & = H(r)^{-1} dx_{||}^2+ dr^2 + r^2 ds^2_{\SS^7} \, , \\
    B_2 & = \big(H(r)^{-1}-1\big) d x^0\wedge d x^1\\
    e^{-2\f} & = e^{-2\f_0} H(r),
\end{aligned} 
\end{equation}
where $\f_0$ is a constant and the harmonic function is given by 
\begin{equation}
    H(r) = 1 + \fr{r_0^6}{r^6}, \quad r_0^6=\fr{(2\p l_s)^6 g_s^2}{6\w_{(7)}}N.
\end{equation}
Performing bi-vector deformation with $\b_2 = \b \dt_0 \wedge \dt_1$ we obtain the following background
\begin{equation} 
\begin{aligned}
    ds^2 & = \left(H(r) - 2\b\right)^{-1} dx_{||}^2+ dr^2 + r^2 ds^2_{\SS^7} \, , \\
    B_2 & = \left(H(r) - 2\b\right)^{-1} d x^0\wedge d x^1\\
    e^{-2\f} & = e^{-2\f_0} \left(H(r) - 2\b\right).
\end{aligned} 
\end{equation}
Now performing a similar rescaling of the longitudinal coordinates as for the deformed Dp-brane backgrounds
\begin{equation}
    x'_{||} = (1-2\b)^{-\fr12}x_{||},
\end{equation}
we arrive at precisely the same string background as in \eqref{eq:F1_initial}, however with a different harmonic function
\begin{equation}
    H(r) = 1 + \fr{r_0^6}{r^6}, \quad r_0^6=\fr{(2\p l_s)^6 g_s^2}{6\w_{(7)}}N', \quad N' = \fr{N}{1-2\b}.
\end{equation}
As before we conclude that such a bi-vector deformation adds a dissolved F1-string charge to the background, thus effectively increasing $N$. When deformation becomes critical ($\b\to 1/2$ in the chosen units) we probe only the near-horizon region of the full string geometry. Inversely, for $\b < 0$ the deformation decreases the string charge mapping the background to the flat space-time in the limit $\b \to - \infty$. Since the deformation is defined in the whole space-time while the charge grows at the core, where gravity is strong, we call this process sedimentation of strings.

Few comments are in place here. First, although it is possible to recover flat space-time from the F1 geometry, the reverse does not seem to be the case. Indeed, the same deformation of the flat space-time after coordinate rescaling gives flat space-time background with the pure gauge Kalb--Ramond field. This means, that although a bi-vector deformation changes NS-NS fields, it not always can be interpreted as adding a dissolved F1-string charge. The conceptual difference between starting with a brane background and the flat space-time seems to be related to the fact that the $N \to 0$ as $\g \to \infty$ but is never actually zero. Assuming, that the deformation adds and remove only dissolved brane charge, might explain why one cannot generate the F1-string background from the flat space-time. 

Second, the calculation above gives a hint for why the same deformation does not reproduce the known thermal Dp-F1 bound states. Indeed, since the deformation transforms the extremal F1 string background into itself with a different value of the charge, it clearly adds only extremal string states to the system. Intuitively, to have a stationary system one should add to a thermal D3-brane with a given temperature string states of the same temperature. Hence, the corresponding deformation must carry information about the blackening function $f(r)$. We were not able to find any such example and thus leave this as an interesting further research direction.

The last comment here concerns the number of strings $N$, that changes continuously upon the deformation. The apparent reason is that we do not impose Dirac charge quantization here and work only with classical supergravity solutions. An interesting question would be how discreteness of the charge appears in deformation of, say, the NS5-brane background.

\textbf{D3}

In \cite{Gubarev:2024tks} a generalization of bi-vector deformations of Type IIB backgrounds was considered. While the standard bi-vector deformations discussed above are elements of the O(10,10) duality group, their generalization given by S-dual bi-vectors and quadri-vectors belong to the U-duality group. Similarly to how the bi-vector $\b^{mn}$ is related to the Kalb--Ramond field $B_{mn}$, its S-dual bi-vector $\b_2{}^{mn}$ and the quadri-vector $\W^{mnkl}$ are related to the R-R forms $C_{mn}$ and $C_{mnkl}$. Therefore, one expects that the S-dual bi-vector must add the D1-brane charge and quadri-vector must add the D3-brane charge. We illustrate that by considering sedimentation of the corresponding Dp-brane backgrounds. 

Start with a quadri-vector deformation of the D3-brane background performed according to the rules listed in Appendix \ref{app:quadri} with
\begin{equation}
    \Omega = \g \, P_0 \wedge P_1 \wedge P_2 \wedge P_3.
\end{equation}
The result can be written as follows
\begin{equation}
\label{eq:D3_deformed}
    \begin{aligned}
        ds'^2 & = \left(H(r)-2\g\right)^{-\fr12} dx_{||}^2 + \left(H(r)-2\g \right)^{\fr12}\left(dr^2 + r^2 d\W_5\right),\\
        C'_{0123} & = -\left(H(r)-2\g\right)^{-1}.
    \end{aligned}
\end{equation}
Since the harmonic function stands also in the block of the metric along transversal coordinates, on has to perform rescaling of both the longitudinal and the radial coordinates:
\begin{equation}
    \begin{aligned}
        x'_{||} & = (1-2\g)^{-\fr14}x_{||},\\
        r' & = (1-2\g)^{-\fr14}r.
    \end{aligned}
\end{equation}
This gives precisely the D3-brane background \eqref{eq:D3_initial} with the same harmonic function. Change in the core charge can be seen as follows. Place a probe string at a distance $r_*$ in the undeformed background. To feel the same force in the deformed background it must sit at a larger distance $r'_* = (1-2\g)^{-1/4}r_* > r_*$. Hence, the core charge is larger.

One may see this explicitly by dropping the rescaling of the radial coordinate and instead rescale the longitudinal coordinates as
\begin{equation}
    \begin{aligned}
        x''_{||} & = (1-2\g)^{-\fr12}x_{||}.
    \end{aligned}
\end{equation}
The resulting metric is equivalent to the metric of the D3-brane up to a constant overall factor, but the harmonic function defined by a rescaled core charge
\begin{equation}
    H = 1 + \fr{1}{r^4}\fr{ g_s (2\p l_s)^{4}}{(7-p)\w_{(5)}}N', \quad N'=\fr{N}{1-2\g}.
\end{equation}
The deformed <<number of branes>> increases for $0\leq\g<1/2$ that again allows to interpret the deformation as adding a (dissolved) D3-brane charge to the background. 

Since as we discuss in Sections \ref{sec:open-closed-D} and \ref{sec:openclosedmembrane}  poly-vectors are related to non-commutativity of open (mem)branes in the same sense as the bi-vector is related to non-commutativity of open strings, the limit $\g \to 1/2$ here is conceptually similar to the critical value of the electric bi-vector deformation when the string becomes non-relativistic. One may speculate that in this limit the D3-branes become non-relativistic, however, since Dp-branes freeze in the string perturbation theory it is not clear how to check this assumption. Instead, one may look at dynamics of the string itself, that in the critical limit $\g \to 1/2$ observes simply the near-horizon geometry of the D3-brane. The near-horizon limit has a simple interpretation that upon increasing the number of branes $N\to \infty$ the dynamics of the probe string at a given distance $r$ is more and more closely described by the near-horizon geometry. Physics behind this is that closed string modes inhabiting the near-horizon region are those of infinitely small energy, such that they cannot escape to  infinity (hence, decouple from the ``bulk'' modes \cite{Aharony:1999ti}). If $N\to \infty$ any energy of the probe string will be small. Therefore, the limit $\g to 1/2$ may be interpreted as moving in the space of string vacua to the sector where massless dynamics gets decoupled (and probably, somehow deformed).

If $\g<0$, the process gets reversed and the deformation removes D3-brane charge from the system. The resulting $N'$ decreases, and as before one cannot generate the D3-brane background by the quadri-vector deformation of the flat space-time.

\textbf{D1}

In the Type IIB theory the standard bi-vector $\b^{mn}$ and the bi-vector $\b_2{}^{mn}$ belong to a single doubled under the SL(2) S-duality group \cite{Gubarev:2024tks}. An S-duality transformation that replaces the axio-dilaton $\t$ by its inverse replaces $\b_2{}^{mn}$ by $\b^{mn}$. Therefore, an S-dual bi-vector deformation by $\b_2 = \b^{mn}\dt_m \wedge \dt_n$ is equivalent to performing S-duality, then the standard bi-vector deformation by $\b = \b^{mn}\dt_\m \wedge \dt_n$ and an S-duality transformation (S$\b$S). 

Given the complexity of the explicit S-dual bi-vector deformations given in \cite{Gubarev:2024tks} it is much more convenient to use the corresponding S$\b$S transformation. Applying it to the D1-brane background we first arrive at the F1 solution \eqref{eq:F1_initial}, its bi-vector deformation has been analysed before, further S-duality gives back the D1-brane background with a rescaled number of branes
\begin{equation}
    N' = \fr{N}{1-2\g}.
\end{equation}
Conclusions here are precisely the same as before and we simply refer to the discussion in the previous sections. 

Such transformation, applied to a Dp-brane background will apparently give a Dp-D1 bound state, that should give the low energy gravitational description in the theory of open D1-branes ending on a Dp-brane. It is natural to expect this

\subsection{Open-closed map for Dp-branes}
\label{sec:open-closed-D}

Let us now return to the interpretation of bi-vector deformations in terms of open strings and the Seiberg--Witten map. Suppose one starts with a background with vanishing B-field and the metric $G_{\m\n}$. This metric is the same for closed strings, that is it stands in the 2d sigma-model action, and for open strings, meaning that it stands in the DBI action. Let us now add the B-field in a very specific way that is (the Seiberg--Witten map)
\begin{equation}
\label{eq:SW_map_1}
    (g+b)^{-1} = G^{-1} + \q,
\end{equation}
where $\q^{mn} = - \q^{nm}$ is a bi-vector. On the LHS we have a family of closed string backgrounds parametrized by $\q$ such that 
\begin{equation}
    \begin{aligned}
        g_{mn}(\q=0) & = G_{mn}, \quad b_{mn}(\q=0) = 0.
    \end{aligned}
\end{equation}
Therefore the map \eqref{eq:SW_map_1} can be understood as deformation of the initial closed string background. 

The fascinating result of \cite{Seiberg:1999vs} is that the DBI action on the background given by $g_{mn}$ and $b_{mn}$
\begin{equation}
\label{eq:DBI_b}
    S_{DBI} = \fr{2\p}{g_s (4\p^2 \a')^{\fr{p+1}{2}}}\int d^{p+1}\x\sqrt{-\det \Big(g+ 2\p \a' (b+F)\Big)},
\end{equation}
were $F=dA$ is the world-volume field strength, is equivalent to the action 
\begin{equation}
\label{eq:DBI_NC1}
    \hat{S}_{DBI} = \fr{2\p}{G_s (4\p^2 \a')^{\fr{p+1}{2}}}\int d^{p+1}\x \sqrt{-\det \Big(G+2\p\a' \hat{F}\Big)},
\end{equation}
where $\hat{F} = d \hat{A} - [\hat{A},\hat{A}]_*$ is the non-commutative field strength. The metric $G_{\m\n}$ is precisely the one given by \eqref{eq:SW_map} and the coupling parameter $G_s$ is defined as
\begin{equation}
    G_s = g_s\sqrt{\fr{\det G}{\det (g+2\p\a' B)}}.
\end{equation}
In the $\a'\to 0$ the action $\hat{S}_{DBI}$ defines non-commutative Yang--Mills theory. Therefore RHS of the map \eqref{eq:SW_map} can be understood as adding non-commutativity to the open string theory. If $\q$ has components along time direction, sending the deformation to its critical value would correspond to decoupling of the NRST sector.

For the case of bi-vector deformations one observes a simple fact: the open string metric does not change. Indeed, upon a bi-vector deformation one only shifts the non-commutativity parameter $\q^{\m\n} \to \q^{\m\n} + \b^{\m\n}$ (if the deformation is defined by the by-vector $\b^{\m\n}$). Demanding the same for a generalization of the non-commutativity parameter to the case of Dp branes in \cite{Berman:2001rka} a generalization of the open/closed map for Dq-branes was suggested
\begin{equation}
\label{eq:berman_map_g}
    G_{mn} = \left( 1 + \frac{1}{(q+1)!} c_{q+1}^{2} \right)^{\frac{1 - q}{1 + q}} \left( g_{mn} + \frac{1}{(q)!} (c_{q+1}^{2})_{mn} \right).
\end{equation}
The generalized non-commutativity parameter is also introduced in \cite{Berman:2001rka} and has the following relation to the closed brane metric and gauge fields
\begin{equation}
\label{eq:berman_map_q}
    \theta^{m_1\ldots m_{q+1}} = - (\a')^{\frac{q+1}{2}} \left( 1 + \frac{1}{(q+1)!} c_{q+1}^{2} \right)^{-\frac{1 - q}{1 + q}} g^{m_1n_1} c_{n_1\ldots n_{q+1}} G^{n_2m_2} \ldots G^{n_{q+1}m_{q+1}}
\end{equation}
These two expressions give the Dp-brane analogue of the RHS of the map \eqref{eq:SW_map} meaning that the theory of an open Dq-brane boundary can be equivalently described in terms of the metric $g_{mn}$ and the gauge field $c_{m_1\dots c_{m_{q+1}}}$ (as in \eqref{eq:DBI_b}) or in terms of the metric $G_{mn}$ and the generalized non-commutativity parameter $\q^{m_1\dots m_{q+1}}$ (as in \eqref{eq:DBI_NC}).

Let us now show that such a relation between open and closed Dp-brane metric follows from poly-vector deformations meaning that these are precisely those deformation under which the open brane metric stays intact. For concreteness we consider the case of D3-branes in Type IIB theory and 4-vector transformations that will also provide expressions relevant to the M2-brane case. Recall the 4-vector transformation rules written in the same notations as the open-closed membrane map
\begin{equation}
\label{eq:quadri_no_C}
    \begin{aligned}
        g_{\m\n} & = K^{\fr12}G_{\m\n}, \\
        g_{mn} & = K^{-\fr12}\left(G_{mn} \pm W_m W_n\right),\\
        c^m & = \pm K^{-\fr14}W^m,
    \end{aligned}
\end{equation}
where the upper/lower sign correspond to the Euclidean/Lorentzian signature of the metric $g_{mn}$ and
\begin{equation}
    \begin{aligned}
        K & = 1\pm W_m W_n G^{mn}, \\
        c^m & - \fr1{4!}\ve^{m n_1\dots n_4}c_{n_1\dots n_4}, \quad W_m =\fr{1}{4!}\ve_{m n_1\dots n_4}\W^{n_1\dots n_4}.
    \end{aligned}
\end{equation}
As in the bi-vector case these 4-vector deformations of a background given by the metric $G_{mn}$ and a vanishing 4-form gauge field can be understood twofold. First, as a family of (closed brane )backgrounds $(g_{mn},c_{m_1\dots m_4})$ parametrised by whatever parameter stands inside $\W^{m_1\dots m_4}$. Second, the background $(G_{mn},\W^{m_1\dots m_4})$ can be understood as a family of open brane metrics and a generalized non-commutativity parameter. 

Now to obtain \eqref{eq:berman_map_g} and \eqref{eq:berman_map_q} one has to go through some calculations. Let us first, expressing $W^m$ in terms of $c^m$ we obtain the following 
\begin{equation}
\label{eq:K}
    K = 1 \pm c^m c^n g_{mn} = 1 + \fr1{4!}c_4^2,
\end{equation}
where $c_4^2 = c_{mnkl}c^{mnkl}$.
Everywhere in these calculations we are using the rule that indices of the gauge field and of the non-commutativity parameter are always raised and lowered by the closed and open metrics respectively. We now proceed with reproducing \eqref{eq:berman_map_g} and write
\begin{equation}
    \begin{aligned}
        G_{mn} & = K^{\fr12}g_{mn}\mp W_m W_n = K^{\fr12}g_{mn}\mp K^{-\fr12}c_m c_n \\
        &=  K^{\fr12}g_{mn} - \fr1{4!}K^{-\fr12}g_{mn}c_4^2 + \fr{1}{3!}K^{-\fr12}c_{m m_1 m_2 m_3}c_m{}^{m_1 m_2 m_3}  \\
        & = K^{-\fr12}\left(g_{mn} + \fr{1}{3!}K^{-\fr12}c_{m m_1 m_2 m_3}c_m{}^{m_1 m_2 m_3}\right),
    \end{aligned}
\end{equation}
where in the last line we used $1/4! c_4^2 = K-1$. Given \eqref{eq:K} this is precisely \eqref{eq:berman_map_g} with $q=3$.

For the generalized non-commutativity parameter we write
\begin{equation}
    \begin{aligned}
        \W^{m_1 \dots m_4} & = \fr{1}{\sqrt{G}}\e^{m_1 \dots m_4 m}G_{mn}c^n K^{\fr14} =  \fr{1}{4!}\fr{1}{\sqrt{G}}\fr{1}{\sqrt{g}}\e^{m_1 \dots m_4 m}\e^{n_1 \dots n_4 n}G_{mn}c_{n_1\dots n_4} K^{\fr14}\\
        &= K G^{m_1 n_1} G^{m_2 n_2}G^{m_3 n_3}G^{m_4 n_4}c_{n_1\dots n_4}.
    \end{aligned}
\end{equation}
Interestingly enough that in this expression any $G^{mn}$ can be replaced by $K^{-\fr12}g^{mn}$, that follows from the simple calculation showing that
\begin{equation}
    W^m c_{mnkl}=0.
\end{equation}
Therefore, performing three such replacement one recovers precisely \eqref{eq:berman_map_q} with $q=3$. We find that performing four replacements produces a nicer expression 
\begin{equation}
    \W^{m_1 \dots m_4} = g^{m_1 n_1} g^{m_2 n_2}g^{m_3 n_3}g^{m_4 n_4}c_{n_1\dots n_4} = c^{m_1\dots m_4},
\end{equation}
where the 4-form of the closed brane background is contracted only with the closed brane metrics. 

Certainly nothing prevents these calculations from being repeated for any other polyvector deformation rules in the Type II supergravity theories. One can also turn on other fields, that will make the formulas incredibly complicated. In any case, the general rule would be that polyvector deformations with vanishing initial gauge fields provide relations between closed and open string/brane backgrounds. We will demonstrate this for 11D backgrounds in Section \ref{sec:openclosedmembrane}

\section{Membranes: OM theory}

Poly-vector deformations of solutions to Type II supergravity equations naturally generalize to 3- and 6-vector deformations of solutions to equations of the 11D supergravity. As we show here explicitly by considering sedimentation of M2 and M5-branes and by generating the M5-M2 bound state tri and six-vector deformations  acquire the same interpretation of adding dissolved M2 and M5-brane charges respectively. This is in consistency with the statement that tri-vector deformations add algebraic structures into the M5-brane theory similar to the non-commutative structure added by (electric) bi-vector deformations into the Dp-brane theory. More details on these algebraic structures can be found in the review \cite{Gubarev:2023jtp} and references therein. Here we illustrate this by demonstrating that the tri-vector deformation formula has the same form as the open-closed membrane map of \cite{Bergshoeff:2000jn} with the tri-vector interpreted as the parameter of non-commutativity of M2 boundaries.

\subsection{Sedimentation of membranes }

\textbf{M2}

Consider the background generated by a stack of $N$ M2-branes
\begin{equation}
\label{eq:M2_initial}
    \begin{aligned}
        ds^2 &= H(r)^{-\fr23}dx_{||}^2 + H(r)^{\fr13}\big(dr^2 + r^2 d\W_{7}\big), \\
        C_{012} &= \pm\big(H(r)^{-1} - 1\big), 
    \end{aligned}
\end{equation}
where
\begin{equation}
    H(r)=1+ \fr{1}{r^6}\fr{N l_P^6}{6\, \w_{(7)}}.
\end{equation}
We now perform a tri-vector deformation of this background using explicit formulas in Appendix \ref{app:tri} with the deformation parameter given by
\begin{equation}
    \Omega = \g \, P_0 \wedge P_1 \wedge P_2 .
\end{equation}
The result can be written as follows
\begin{equation}
\label{eq:M2_deformed}
    \begin{aligned}
        ds'^2 & = \left(H(r)-2\g\right)^{-\fr23} dx_{||}^2 + \left(H(r)-2\g \right)^{\fr13}\left(dr^2 + r^2 d\W_5\right),\\
        C'_{012} & = \pm\left(H(r)-2\g\right)^{-1},
    \end{aligned}
\end{equation}
where as in the Type II case we observe the critical value of the deformation parameter. 

As before it is possible to show that the obtained background corresponds to a background of a stack of $N'$ M2 branes in two ways. One is to rescale both the longitudinal coordinates and the radial coordinate as
\begin{equation}
    \begin{aligned}
        x'_{||} & = (1-2\g)^{-\fr13}x_{||}, \\
        r' & = (1-2\g)^{\fr16}r.
    \end{aligned}
\end{equation}
In this case one obtains precisely the background \eqref{eq:M2_initial} however the probe M2-brane has to be placed further from the core to experience the same force, that reflects the increased number of branes at the core. To recover the precise relation between $N'$ and $N$ we rescale only the longitudinal coordinates as
\begin{equation}
    \tilde{x}_{||} = (1-2\g)^{-\fr12}x_{||},
\end{equation}
that gives
\begin{equation}
\label{eq:M2_rescaled}
    \begin{aligned}
        (1-2\g)^{-\fr13}ds'^2 & = \tilde{H}^{-\fr23} d\tilde{x}_{||}^2 + \tilde{H}^{\fr13}\left(dr^2 + r^2 d\W_5\right),\\
        (1-2\g)^{-\fr12}C' & = \pm\tilde{H}^{-1},
    \end{aligned}
\end{equation}
where components of the 3-form are written in the $\tilde{x}_{||}$ coordinates. The new defining function is given by
\begin{equation}
    \tilde{H}(r) = 1 + \fr1{r^6}\fr{N' l_P^6}{6\, \w_{(7)}}, \quad N'=\fr{N}{1-2\g}.
\end{equation}
As expected, we observe that the initial number of branes increases with the deformation supporting the interpretation that tri-vector deformation adds dissolved M2-brane charge.

Close to the critical value $\g \sim 1/2$ the M2 background asymptotes its near-horizon limit that is $\AdS_4\times\SS^7$. Conversely, one may start with the near-horizon limit and recover the full M2-brane background. In this case the deformation will remove extra M2-brane charge and restore the correct asymptotics of the solution (upon the proper rescaling).

\textbf{M5}

To sediment dissolved M5-brane charge one should apply 6-vector deformations to the background created by a stack of $N$ M5-branes, that is
\begin{equation}
\label{eq:M5_initial}
    \begin{aligned}
        ds^2 & = H(r)^{-\fr13} dx_{||}^2 + H(r)^{\fr23}(dr^2+ r^2 d\W_4), \\
        C_{01\dots 5} & = (H(r)^{-1} - 1),
    \end{aligned}
\end{equation}
where $C_{01\dots 5}$ is the gauge potential magnetically dual to the 3-form and the defining function reads
\begin{equation}
    H(r) = 1 + \fr{1}{r^3}\fr{N l_P^3}{3 \w_{(4)}}.
\end{equation}
In this case no explicit expressions for deformation rules are available at this point. Indeed, the formalism of \cite{Gubarev:2020ydf} is valid only for backgrounds with the metric of the block $5+6$ form and no 3-form field in the 5d subspace, that is clearly not the case. To override that we perform 6-vector deformations as defined in \cite{Gubarev:2020ydf} in the full $\rmE_{6(6)}$ exceptional field theory of \cite{Hohm:2013vpa}, essentially using a formalism with external fluxes similar to \cite{Barakin:2024rnz}. The six-vector deformation is taken as
\begin{equation}
    \W_6 = \g\,  P_0 \wedge P_1 \wedge P_2 \wedge P_3\wedge P_4 \wedge P_5,
\end{equation}
and the resulting background reads
\begin{equation}
\label{eq:M5_deformed}
    \begin{aligned}
        ds'^2 & = \left(H(r)-2\g\right)^{-\fr13} dx_{||}^2 + \left(H(r)-2\g \right)^{\fr23}\left(dr^2 + r^2 d\W_5\right),\\
        C'_{01\dots 5} & = \left(H(r)-2\g\right)^{-1},
    \end{aligned}
\end{equation}
As before, rescaling both the longitudinal and radial coordinates as
\begin{equation}
    \begin{aligned}
        x'_{||} & = (1-2\g)^{-\fr16}x_{||},\\
        r' & =  (1-2\g)^{\fr13}r
    \end{aligned}
\end{equation}
one obtains again the M5-brane background but the M2-probe experience the same force further from the core. Rescaling only the longitudinal coordinate as
\begin{equation}
    \tilde{x}_{||}  = (1-2\g)^{-1}x_{||}
\end{equation}
we obtain the background
\begin{equation}
    \begin{aligned}
        (1-2\g)^{-\fr23}ds'^2 & = \tilde{H}(r)^{-\fr13} d\tilde{x}_{||}^2 + \tilde{H}(r)^{\fr23}(dr^2+ r^2 d\W_4), \\
        (1-2\g)^{-1}C_{01\dots 5} & = (\tilde{H}(r)^{-1} - 1),
    \end{aligned}
\end{equation}
where $\tilde{H}(r)$ is the standard defining function with 
\begin{equation}
    N' = \fr{N}{1-2\g}.
\end{equation}
Again we see that the deformation increases the number of M5-branes at the core of the solution. Close to the critical value $\g \sim 1/2$ the background can be approximated by the near-horizon geometry of the M5-brane that is $\AdS_7\times \SS^4$. Similarly to all previous cases the full M5-brane geometry can be restored from the near-horizon approximation by the same 6-vector deformation. The interpretation would be  that of removing the M5-brane charge.

\subsection{M5-M2 bound state}

Given the above discussion, one expects that tri-vector deformations of the M5-brane generate M5-M2 bound state adding dissolved M2-brane charge. Due to the complications with thermal states, that arise here in the same form as for the Dp-F1 bound state, we restrict ourselves to only extremal backgrounds. Because of the flux configuration of the initial M5-brane solution one must use the formalism of deformations with external fluxes that for tri-vector deformations has been constructed in \cite{Barakin:2024rnz}. 

Starting with \eqref{eq:M5_initial} and taking 
\begin{equation}
    \W = \g \dt_0 \wedge \dt_1 \wedge \dt_2,
\end{equation}
we obtain the following deformed background
\begin{equation} 
\label{eq:M5-M2}
\begin{aligned}
    ds'{}^2  = &\ H_0^{\fr13} \left(H_0 -  \g^2\right)^{-\fr23}\left(-(dy^0)^2+(dy^1)^2+(dy^2)^2\right) \\
    &+ H_0^{-\fr23} \left(H_0 -  \g^2\right)^{\fr13} \left((dy^3)^2+(dy^4)^2+(dy^5)^2\right)  + H_0^{\fr13} \left(H_0 -  \g^2\right)^{\fr13} ( dr^2 + r^2 d\W_4) , \\ 
    F'_4  =& \fr{N l_P^3}{\w_{(4)}} d\vol[\SS^4] - \g d \left( H_0^{-1}  dy^3\wedge dy^4 \wedge dy^5  -  (H_0 -  \g^2)^{-1} dy^0\wedge dy^1\wedge dy^2  \right),
\end{aligned} 
\end{equation}
where we denoted by $H_0$ the initial defining function, and $y^0,\dots,y^5$ are the longitudinal coordinates to be rescaled. Denoting $\g=\sin \q$ we perform the following rescaling
\begin{equation}
\begin{aligned}
    y^{0,1,2} &= \cos^{\frac{2}{3}} \q  \, x^{0,1,2}, \\ y^{3,4,5} &= \cos^{-\frac{1}{3}} \q  \, x^{3,4,5}, \\
    r & = \cos^{-\frac{1}{3}} \q  \, \r,
\end{aligned}
\end{equation}
and define
\begin{equation}
\begin{aligned}
     H  & = \fr{1}{\cos^2\theta}\left(H_0 - \sin^2\theta\right) = 1 + \fr{N}{\cos \q}\fr{l_P^3}{3\w_{(4)}}\fr{1}{\r^3}, \\
     D^{-1} & = \cos^2\theta + \sin^2\theta H^{-1}.
\end{aligned}
\end{equation}
The solution then takes the following form
\begin{equation} 
\begin{aligned}
    ds'{}^2  = &\ H^{-\fr13}D^{-\fr13} \left(-(dx^0)^2+(dx^1)^2+(dx^2)^2\right) \\
    &+ H^{-\fr13}D^{\fr23}\left( (dx^3)^2+(dx^4)^2+(dx^5)^2\right)  + H^{\fr23}D^{-\fr13} ( d\r^2 + \r^2 d\W_4) ,\\
    F'_4  =& \fr{N l_P^3}{\w_{(4)}} d\vol[\SS^4] - \sin \q d_{\rho} \left( D H^{-1} \cos^{-1} \q dx^3\wedge dx^4 \wedge dx^5  -  H^{-1} dx^0\wedge dx^1\wedge dx^2  \right)
\end{aligned} 
\end{equation}
or in components
\begin{eqnarray}
F_{012\r} &=& - \sin {\theta}  \dt_\r ({H}^{-1})
\\
F_{345\r} &=& \tan {\theta} \dt_\r ({D} {H}^{-1})
\\
F_{7\,8\,9\,10} &=& \fr{N l_P^3}{\w_{(4)}} \sin^3(x_8)  \sin^2(x_9)  \sin (x_{10}) 
\end{eqnarray}
This has the same form as in \cite{Harmark:2000ff} (with the misprint in $F_{7\,8\,9\,10}$ corrected) and we conclude that such a tri-vector deformation indeed generates the background of the M5-M2 bound state. 

It is interesting to notice, that in contrast to the Dp-F1 bound states one is required to rescale not only the M2-charge directions, but also the directions along the M5-brane and the radial direction. It is tempting to relate such a difference to the fact that M-theory does not have a natural limit (like $g_s \to 0$) under which M5 and M2 branes behave essentially differently.

\subsection{Membrane open/closed map}
\label{sec:openclosedmembrane}

Our final stop is to show that the formula of tri-vector deformations is the same as the open-closed membrane map as it has been found in \cite{Bergshoeff:2000jn}. The open membrane metric and non-commutativity parameter read
\begin{equation}
\label{eq:open_closed_membrane}
    \begin{aligned}
    G_{mn} & = \left( 1 + \frac{1}{3!} c_{3}^{2} \right)^{\frac{1}{3}} \left( g_{mn} + \frac{1}{2!} (c_{3}^{2})_{mn} \right), \\
    \theta^{mnk} & = -  \left( 1 + \frac{1}{3!} c_{3}^{2} \right) c^{mnk} .
    \end{aligned}
\end{equation}
Here $(c_{3}^{2})_{mn} = c_{mpq} c_{nrs} g^{pr} g^{qs}$. For the initial background with vanishing 3-form gauge field the tri-vector transformations give
\begin{equation}\label{Cdef}
\begin{aligned}
        g_{\m\n} &= K^{-\fr13}G_{\m\n}, \\
        g_{m n} & = K^{\fr23}\left( G_{mn} \pm W_m W_n \right), \\
     	c^{mnk} & =\mp K^{-1}\W^{mnk}, \\
        K^{-1} & =  1 \pm   W_m W^m,
\end{aligned}
\end{equation}
where $W_m = 1/3! \varepsilon_{mnkl}\W^{nkl}$ and upper/lower signs correspond to Euclidean/Minkowskian signatures of $g_{mn}$ (the block $g_{\m\n}$ is always of the opposite signature such that the full space-time has one time direction). Due to the huge similarities between these transformations and the quadri-vector deformations formula \eqref{eq:quadri_no_C} reversing the expressions \eqref{eq:open_closed_membrane} goes pretty much along the same lines as the derivation in Section \ref{sec:open-closed-D}. Note however that  \ref{eq:open_closed_membrane} is not the same as \eqref{eq:berman_map_g} and \eqref{eq:berman_map_q} with $q=2$.

Taking determinant of the second line in \eqref{Cdef} we recover $\det g = K^{5/3}\det G$, that allows to write
\begin{equation}
    v_{m} = \pm K^{-\fr16}W_m.
\end{equation}
Here $v^m = 1/3! \varepsilon^{mnkl}c_{nkl}$ and the indices on the LHS and RHS are raised and lowered by $g_{mn}$ and $G_{mn}$ respectively. Using this relation we obtain
\begin{equation}
    K = 1 \pm v^2 = 1 + \fr1{3!}c_3^2
\end{equation}
and finally
\begin{equation}
    \begin{aligned}
        G_{mn} & = \left( 1 + \frac{1}{3!} c_{3}^{2} \right)^{\frac{1}{3}} \left( g_{mn} + \frac{1}{2!} (c_{3}^{2})_{mn} \right), \\
        \W^{mnk} & = -  \left( 1 + \frac{1}{3!} c_{3}^{2} \right) c^{mnk},\\
        G_{\m\n} & = \left( 1 + \frac{1}{3!} c_{3}^{2} \right)^{\frac{1}{3}} g_{\m\n}.
    \end{aligned}
\end{equation}
This is precisely the open-closed membrane map with the non-commutativity parameter $\q^{mnk}$ replace by the tri-vector deformation parameter $\W^{mnk}$, and endowed with the relation for the $7\times 7$ blocks $G_{\m\n}$ and $g_{\m\n}$ of the metrics.

\section{Conclusions and discussions}

In this work we investigate transformation of extremal brane backgrounds of Type II and $D=11$ supergravities under poly-vector deformations and interpret the result in terms of brane states. We find that the standard bi-vector deformations applied to Dp-brane backgrounds (we analyse explicitly D3 and D1) generate extremal Dp-F1 bound state backgrounds. In \cite{Harmark:2000wv} their critical limit has been conjectured to provide the holographic description of non-relativistic string theory (NRST). More accurately speaking, such bound state backgrounds are the NRST analogues of Dp-brane background in string theory. This result collocates nicely with the interpretation of the NCOS theory (the open string sector of NRST) as (the critical limit) of the open string theory on a background with non-vanishing space-time components of the Kalb--Ramond field. Indeed, the bi-vector deformation \eqref{eq:NS_transf} has precisely the same form as the open-closed string map of \cite{Seiberg:1999vs} and hence can be understood as turning on the space-time non-commutativity.

We extend this picture to all other known deformations: the S-dual bi-vector and quadri-vector deformations in the Type IIB theory, and tri- and six-vector deformations in the 11D theory. We find that in all these cases poly-vector deformations of a (mem)brane background generate bound states of the initial (mem)brane with the corresponding dissolved BPS charge. The charges are that of the D1- and D3-branes in Type IIB and of the M2 and M5 branes in 11D. The claim is supported by several observations. First, the resulting background reproduce those of the bound states considered in \cite{Harmark:2000wv,Harmark:2000ff} (up to a few misprints we have found in the text there). Second, we find that such deformations applied to a stack of branes of the same charge (e.g. quadri-vector to the D3 background) change the number of branes at the core. We call this process sedimentation since the deformation formally acts in the whole space-time, while the charge increases at the core.

Similarly to how Dp-F1 bound state backgrounds are natural brane description in NRST, M5-M2 bound states describe branes in the OM (open membrane) theory of \cite{Gopakumar:2000ep}. This has been discussed previously in \cite{Harmark:2000ff} and in this work we develop it further  by connecting to the open-closed membrane map of \cite{Bergshoeff:2000jn}. Concretely we find that tri-vector deformations can also be interpreted as transformation, that turn on the tri-vector non-commutative parameter. The same is true for the Type IIB theory where the quadri-vector deformations reproduce the open-closed Dp-brane map.

The perspective we develop here is far from being complete as many questions remain unanswered. To start with we return to the issue of thermal backgrounds, that cannot be obtained by a TsT of a non-extremal brane solution. The most natural explanation seems to be that such deformations only add extremal states and hence cannot generate thermal states in equilibrium. An interesting question would be to analyse the deformation, that connect thermal bound states and non-extremal brane solution. A related question concerns the possible interpretation of non-abelian deformations in a similar way. For example, a deformation of the type $P\wedge M$ (momentum and angular momentum) could correspond to adding a rotating string charge. We hope to report soon on a progress in this direction. 

It is worth to mention a subtle point related to changing of the $r_0$ parameter in the harmonic function related to the core charge of the Dp-brane. On one hand this can be interpreted as changing the number of Dp-branes on which the strings end. On the other hand the Seiberg--Witten map is formally valid only for a single Dp-brane or for the U(1) sector of the $\rmU(N)$ strings ending on a stack of N Dp-brane. A generalization of the map to the $\rmU(N)$ case has been suggested in \cite{Ma:2020msx} and analysed further in \cite{Ma:2023hgi,Ma:2025rti} (see \cite{Ma:2025zaz} for a review).  It would interesting to clarify this subtlety and analyse to what extent one is able to interpret changing in $r_0$ as changing in the number of branes.

The S-dual bi-vector transformation, applied to a Dp-brane background will apparently give a Dp-D1 bound state, that should give the low energy gravitational description in the theory of open D1-branes ending on a Dp-brane. It is natural to expect this theory to be electro-magnetically dual to the NC YM theory that is the weak field description of the Dp-F1 bound state. Such duality relations were considered in  \cite{Ho:2015mfa} and in \cite{Russo:2000mg} and it would be interesting to analyse these relations further.

Another large topic that we didn't touch in this work is that of uni-vector deformations in the dilaton Einstein--Maxwell theory developed in \cite{Gubarev:2025hvr}. In general one would expect a similar behaviour, however, such a theory does not have much variety of extended objects. An interesting question would be whether a uni-vector deformation of a magnetically charged black hole generates a dyonic state. A closer investigation of uni-vector deformations has a huge advantage that these transformations are the usual coordinate transformations in the parent theory given by the Einstein-Hilbert action in one dimension larger. This aligns with a different viewpoint provided by the formalism of exceptional field theories, where T(U)-duality symmetries are made covariant after extending the space-time by extra <<dual>> coordinate. Bi-vector deformations appear to be a certain class of coordinate transformations in this extended space (see \cite{Sakamoto:2017cpu,Gubarev:2025hvr}). On the other hand in \cite{Bakhmatov:2016kfn,Bakhmatov:2017les,Berman:2018okd} it has been show that various branes of string theory can be understood as projection of different orientations of some generalized brane object to the normal geometric space. Combining these results with the observations here one concludes that Dp-F1, Dp-D1, M5-M2 and similar bound states can also be interpreted as projection of the initial brane after a generalized coordinate transformation.

\section*{Acknowledgements}

The authors would like to thank David Berman, Riccardo Borsato, Shahin Sheikh-Jabbari and Stijn van Tongeren for valuable discussions and useful suggestions.


\appendix 

\section{Type II D=10 and D=11 supergravities}

Equations of motion for the Type II supergravity read
\begin{equation}
    \begin{aligned}    
        R_{mn} - \fr14 H_{mpq}H_n{}^{pq}  +2 \nabla_{m}\nabla_{n}\f & = T_{mn} ,\\
        -\fr12 \nabla^k H_{kmn} +\nabla^k \f H_{kmn}  & = K_{mn},\\ 
        R - \fr12 |H_3|^2 + 4 (\Box \f -\nabla^m\f \nabla_m\f) & = 0,\\
        d *\mF_p - H_3 \wedge * \mF_{p+2}  &= 0,
    \end{aligned}
    \label{eq:gensugra0}
\end{equation}
where
\begin{equation}
    \begin{aligned}    
        T_{mn} & =\fr14 e^{2\Phi}\sum_p\left[\fr1{p!}\mF_m{}^{k_1\dots k_{p}}\mF_{nk_1\dots k_{p}} - \fr12 g_{mn}|\mF_{p+1}|^2\right], \\
        K_{mn}   &= \fr14 e^{2\Phi}\sum_p\fr1{p!}\mF_{k_1\dots k_p}\mF_{mn}{}^{k_1\dots k_p}.
    \end{aligned}
\end{equation}
We denote $|\w_p|^2 = \frac{1}{p!} \w_{i_1\dots i_p}\w^{i_1\dots i_p}$ for a $p$-form $\w_p$.

For our solutions to 11-dimensional supergravity equations we follow the conventions of \cite{Ortin:2015hya}. The action is given by
\begin{equation}
    S_{11} = \k \int d^{11} x \sqrt{-g} \left(R - \fr{1}{2 \cdot 4!}F_{\m\n\r\s}F^{\m\n\r\s} - \fr{1}{(144)^2} \fr{1}{\sqrt{-g}}\e^{\m_1\dots \m_{11}}F_{\m_1 \dots \m_4}F_{\m_5\dots \m_8}C_{\m_9\m_{10}\m_{11} }\right),
\end{equation}
with $F_{\m\n\r\s} = 4 \dt_{[\m}C_{\n\r\s]}$. Equations of motion then read
\begin{equation}
    R_{\m\n} - \fr12 R g_{\m\n} - \fr{1}{2\cdot 3!} F_{\m\r\s\l}F_\n{}^{\r\s\l} + \fr{1}{2}\fr{1}{2\cdot 4!} F_{\m_1\dots \m_4}F^{\m_1\dots \m_4}g_{\m\n}=0,
\end{equation}
for the metric and
\begin{equation}
    \dt_{\m}\left(\sqrt{-g}F^{\m\n\r\s}\right) - \fr{1}{8\cdot 4! \cdot 3!} \e^{\n\r\s\m_1\dots \m_8}F_{\m_1\dots \m_4}F_{\m_5\dots \m_8}=0.
\end{equation}
for the 3-form gauge field.

\section{Bi-vector deformations}
\label{app:bi}

Under a bi-vector deformation generated by $\b_2 = \b\, \dt_0 \wedge \dt_1$ the NS-NS sector of Type IIA/B supergravity transforms according to
\begin{equation}
\label{eq:NS_transf}
    \begin{aligned}
        g^{-1}+ \b &= (G+B)^{-1},\\
        e^{2\f} \sqrt{\det g} & = e^{2\Phi}\sqrt{\det  G}.
    \end{aligned}
\end{equation}
Here the matrices $g$ and $G$ stand for the initial and the deformed metrics respectively (in the string frame), the matrix $B$ is the deformed Kalb--Ramond field and $\Phi$ is the deformed dilaton.  

Transformation of the fields in the R-R sector can be most conveniently written in terms of the gauge invariant field strengths defined as
\begin{equation}
    \mc{F}_p = F_p + H_3 \wedge C_{p-3},
\end{equation}
where $F_p = dC_{p-1}$. Defining the formal sum of all such forms as $\mF$ we can write the transformation as follows
\begin{equation}
    \mF' \wedge e^{-B} = (1+ \iota_{\b})\mc{F}\wedge e^{-b},
\end{equation}
where $\iota_{01}$ denotes inner product with $\b_2$ (note the sign difference in the exponent w.r.t. \cite{Imeroni:2008cr}).

In components for the Type IIB case, that is of relevance for the main text of the paper we have
\begin{equation}
\label{eq:RR_transf}
    \begin{aligned}
        \mF'_1& = \mF_1 + \iota_\b\left(\mF_3 - \mF_1 \wedge b\right),\\
         \mF'_3 - \mF'_1 \wedge B& = \mF_3 - \mF_1 \wedge b + \iota_\b\left(\mF_5 - \mF_3 \wedge b + \fr12 \mF_1 \wedge b^2\right),\\
          \mF'_5 - \mF'_3 \wedge B + \fr12 \mF'_1 \wedge B^2 & = \mF_5 - \mF_3 \wedge b + \fr12 \mF_1 \wedge b^2  \\
          &\quad +\iota_\b\left(\mF_7 - \mF_5 \wedge b + \fr12 \mF_3 \wedge b^2- \fr16 \mF_1 \wedge b^3\right).
    \end{aligned}
\end{equation}

\section{Quadri-vector deformation in Type IIB}
\label{app:quadri}

Quadri-vector deformations, that appear in Type IIB theory as a part of the U-duality group $\rmE_{6(6)}$, have been developed in \cite{Gubarev:2024tks} for backgrounds of  a particular form
\begin{equation}
    \label{eq:bg10truncated}
    \begin{aligned}
        ds_{10}^2 & = g^{-\fr13}g_{\m\n}(y,x)dy^\m dy^\n + g_{mn}(x)dx^m dx^n ,\\
        C_4 & = \fr1{4!}C_{\m_1\dots \m_4}dy^{\m_1}\wedge \dots \wedge dy^{\m_4} + \fr1{4!}C_{m_1\dots m_4}dx^{m_1}\wedge \dots \wedge dx^{m_4}.
    \end{aligned}
\end{equation}
On top of this the 5-form field strength is required to be self-dual in ten dimensions. Deformed backgrounds will also follow the same ansatz and the deformation rules read
\begin{equation}
\begin{aligned}
   g'_{\mu \nu}&=K^{\frac{1}{2}} g_{\mu \nu}\\
    c'{}^{m }&=K^{-\frac{1}{4}} c^{m }\left(1+ W  c \right) \pm K^{-\frac{1}{4}}W^m  \\
    g'_{m n} &= K^{-\fr12} \left(g_{m n}
        +2\, W_{(m} c_{n)} \pm \left(1\pm c^2\right) W_m W_n \right),\\
    W_m&=\frac{1}{4!}\varepsilon_{m p_1 \ldots p_4} \W^{ p_1 \ldots p_4}, \\
    g' & = K^{-\fr32}g,
\label{deformed_IIB_fields}
\end{aligned}
\end{equation}
where indices on the RHS are raised and lowered by the undeformed metric $g_{mn}$, and the upper(lower) sign corresponds to Euclidean(Minkowskian) signature of the metric $g_{mn}$.
\begin{equation}
    \begin{aligned}
        c^m & - \fr1{4!}\ve^{m n_1\dots n_4}c_{n_1\dots n_4}, \quad W_m =\fr{1}{4!}\ve_{m n_1\dots n_4}\W^{n_1\dots n_4},\\
       K&= \left(1+W c\right)^2 \pm W^2
    \end{aligned}
\end{equation}
Transformation rules for the internal and external components of the self-dual 5-form field strength can be derived by simply taking a covariant derivative of $c^m$. For the internal part we have
\begin{equation}
    \label{eq:deform_int}
    F_{int}' = \fr{1}{5!} \dt_{m}\left(K^{-\fr{3}{4}}g^{\fr12}c'{}^m\right) \e_{n_1\dots n_5} dx^{n_1}\wedge\dots \wedge dx^{n_5},
\end{equation}
and the self-duality condition gives the external components
\begin{equation}
    \label{eq:deform_ext}
    F_{ext}' = K^2g^{-\fr12}\dt_m\left(K^{-\fr{3}{4}}g^{\fr12}c'{}^m\right) \fr{1}{5!}\sqrt{-\det g_{\m\n}}\e_{\m_1 \dots \m_5}dx^{\m_1}\wedge \dots \wedge dx^{\m_5}.
\end{equation}
Note using both $g$ and $g'$ in the expression

\section{Tri-vector deformations in D=11 supergravity}
\label{app:tri}

Tri-vector transformation formulas have been presented in \cite{Barakin:2024rnz} for backgrounds of the form
\begin{equation}
    \begin{aligned}
        ds_{11}^2 & = g^{-\fr15} g_{\m\n}dx^\m dx^\n + g_{mn} \big(dx^m + A_\m{}^m dx^\m\big)\big(dx^n + A_\n{}^n dx^\n\big), \\
        \hat{F}_{(4)} & = \fr{1}{4!} F_{\hat{\m}\hat{\n}\hat{\r}\hat{\s}} dx^{\hat{\m}}\wedge dx^{\hat{\n}}\wedge dx^{\hat{\r}}\wedge dx^{\hat{\s}} \\
        & = F_{(4)} + F_{(3) m} \wedge dx^m + \fr12 F_{(2) mn}dx^{mn} + \fr1{3!} F_{(1)mnk}dx^{mnk}+ \fr1{4!}F_{mnkl}dx^{mnkl},
    \end{aligned}
\end{equation}
where $g = -\det g_{mn}$. We also use the notations $dx^{m_1\dots m_p} = dx^{m_1}\wedge \dots \wedge dx^{m_p}$ and $F_{(p)m_1\dots m_k}$ is a $p$-form in the external 7-dimensional space-time with $k$ indices. 

As before we introduce the following short-hand notations
\begin{align}
    K = & ( 1 - W_{k} V^{k} )^2 - W^{2} \, , \text{ where}\\
    W_m = \fr1{3!} g^{\fr12} \, \e_{mnkl} & \W^{nkl} \quad \text{and} \quad V^{m} = \fr1{3!} \, g^{-\fr12} \, \e^{mnkl} A_{nkl},
\end{align}
that allow to write transformation rules in the following form
\begin{equation}
\label{eq:transf_gauge}
    \begin{aligned}
         F'_{\mu \nu}{}^m & = F_{\mu \nu}{}^m - \fr12 \W^{m k l} ( F_{\mu \nu k l} - F_{\mu \nu}{}^n A_{n k l} ) \\
         F'_{m n k l} & = 4 \dt_{[m} \left[ K^{-1} A_{n k l]} \, ( 1 - W_{k} V^{k} ) - \w^6 \W_{n k l]}\right] \\
         F'_{\mu m n k} & = D_\mu \left[ K^{-1} A_{m n k} \, ( 1 - W_{k} V^{k} ) - K^{-1} \W_{m n k} \right] + 3 (K^{-1} - 1) \dt_{[m} A_{|\mu| n k]} \\
         F'_{\mu \nu m n}& = F_{\mu \nu m n} - F_{\mu \nu}{}^{k} A_{m n k} \\
         &\quad + K^{-1} \left[F_{\m\n}{}^k - \fr12 \W^{kpq} \left(F_{\m\n pq} - F_{\m\n}{}^lA_{pql}\right)\right] \cdot \left[A_{mnk} ( 1 - W_{k} V^{k} ) - \W_{mnk}\right] ,\\
         F'_{\mu \nu \rho m} & = F_{\mu \nu \rho m} + \fr12 \W^{k l n} ( F_{[\mu \nu |k l|} - F_{[\mu \nu}{}^p A_{p k l} ) A_{\rho] n m} \\
         & - W_m \left( \fr{e}{4!} \epsilon_{\mu \nu \rho \sigma \xi \eta \zeta} F^{\sigma \xi \eta \zeta} g^{\fr1{10}} + V^n ( F_{\mu \nu \rho n} + F_{[\mu \nu}{}^k A_{\rho] k n} ) \right) \\ 
         F'_{\mu \nu \rho \sigma}& = ( 1 - W_m V^m ) F_{\mu \nu \rho \sigma} - W^m \fr{e}{3!} \epsilon_{\mu \nu \rho \sigma \xi \eta \zeta} ( F^{\xi \eta \zeta}{}_m + F^{[\xi \eta |n|} A^{\zeta]}{}_{n m} ) g^{-\fr1{10}},
\end{aligned}\end{equation}
where $e^2 = \det g_{\m\n}$.

\addcontentsline{toc}{section}{References}
\bibliography{bib.bib}
\bibliographystyle{utphys}

\end{document}